# Leveraging Multiple Online Sources for Accurate Income Verification

Chirag Mahapatra[1] and Kedar Bellare[1]

[1] Airbnb, San Francisco CA 94103, USA

**Abstract.** Income verification is the problem of validating a person's stated income given basic identity information such as name, location, job title and employer. It is widely used in the context of mortgage lending, rental applications and other financial risk models. However, the current processes surrounding verification involve significant human effort and document gathering which can be both time-consuming and expensive. In this paper, we propose a novel model for verifying an individual's income given very limited identity information typically available in loan applications. Our model is a combination of a deep neural network and hand-engineered features. The hand-engineered features are based upon matching the input information against income records extracted automatically from various publicly available online sources (e.g. payscale.com, H-1B filings, government employee salaries). We conduct experiments on two data sets, one simulated from H-1B records and the other from a real-world data set of peer-to-peer (P2P) loan applications obtained from one of the world's largest P2P lending platform. Our results show a significant reduction in error of 3-6% relative to several strong baselines. We also perform ablation studies to demonstrate that a combined model is indeed necessary to achieve state-of-the-art performance on this task.

## 1. Introduction

In this paper, we address the problem of income verification: given a person's basic identity attributes (e.g. name, date of birth), current employment information (e.g. job title, company, location), and stated income we attempt to accurately validate the income of the given individual within a certain threshold. We define income as sum of earnings such as base salary and bonus in a year. We do not include rent, stocks, interest payments, dividend payments and other forms of income in our modeling. We would like to note that we use identity attributes such as name only to look up relevant income records via database queries or web searches. We do not use names for modeling to avoid .

One of the foremost use-cases of income verification is to identify creditworthy and fraudulent users during loan applications. Several recent pay-day loan and peerto-peer loan companies provide instant/pre-approved loans, powered by models behind the scenes that can accurately assess the risk involved with an application. An individual's validated income is an important feature in these risk models. Banks and lending institutions need to make these checks in a cost-effective and time-efficient manner. Currently applicants need to provide documents which have to be manually verified which can be both time-consuming and expensive. We present an approach that can algorithmically validate the input information quickly and accurately.

The main component of an income verification system is income prediction from a given input identity. We leverage the power of publicly available data sources on the web to solve the prediction problem. Obviously, using the public web comes with the following challenges:

- **Search, Extract, and Match:** We need to build queries from input to get a candidate set of web documents and database records that contain income information. We then have to extract data from structured and unstructured sources and filter the sources to keep only those which have the closest match to the input identity.

- **Partial and ambiguous information in the input and the Web:** An example of this is acronyms and alternate names of companies. For example, "United States Postal Service" can be represented as "USPS", "U.S.P.S", "US Postal Service" etc. We need to be able to identify that these are all the same entity.

- **Erroneous data on the Web:** Not all sources on the web are accurate. Some web sources misrepresent salary information or are out dated.

We have built a robust system that addresses all the aforementioned issues. First, our algorithm leverages search engines to surface salary content on tail entities; the surfaced content includes domains such as payscale.com, salary.com, H1-B data, etc. We have also crawled a large number of public government databases and indexed the available information in a structured format so that they can be searched directly. Each of these domains pose a different extraction challenge. For example, in some cases we need to extract content from unstructured text, whereas in other cases we can directly use wrappers based on XPath expressions tailored to specific websites.

Second, to address the issue of partial information (either in the input or on the Web), we often "expand" the scope of our identity: for example, we infer the industry from the company, the experience/level from job title and date of birth if present, and then generalize our search to the given position and industry.

Finally, to address the issue of possibly incorrect information on the Web, we build a model that aggregates salary ranges across several domains, and then computes one unified range by factoring in (a) frequencies of occurrence, (b) trustworthiness of sources, and (c) strength of identity match between the input and a source. For example, if multiple sources indicate that the median salary of a software engineer in San Francisco is around $100,000, but one source indicates the median salary is $50,000, the algorithm would center the range around $100,000.

**Table 1.** Example Input

| Attribute | Example |
| --- | --- |
| Name | Barack H Obama |
| Address | Washington DC |
| Date of Birth | August 4, 1961 |
| Employer | United States Government |
| Position | President |
| Stated Income | $400,000 |

We have made the following assumptions in our work:
- Employment information is available and correct.

- Identity information is available and correct.
- Verified income provided by the client does not include rent, interest payment, dividends, debts etc.

We introduce the model and delve into each component in detail in Section 2. We discuss the experiments used to validate our model in Section 3. We cover the previous literature in income modeling in Section 4 and conclude the paper in Section 5.

## 2.    Method

In this section, we first describe the input provided for income verification (Section 2.1) and outline our overall approach to tackling it (Section 2.2). Then in Section 2.4 we delve into the system design and describe the individual system components at a high-level. Sections 2.3 and 2.5 lay out the crux of this paper by describing the systems utilizing online sources and input data respectively in detail.

### 2.1.    Input Description

Table 1 gives an example input to our system. Each row in the input has personally identifiable information (e.g. name, address, dob), employment information (e.g. employer, job title) and the individual's stated income. Our goal in this paper is to verify that the stated income is accurate. There are two things to note about the input information here.

- Firstly, some of the information may be missing or incomplete. For example, the second row in the table has an incomplete address. However, we are always provided with the employment and income information although in some cases this information may be noisy or incomplete. In this paper, we focus on the problem of verifying the income assuming that the employment information is correct. In practice, we have observed that a very small percentage (i.e. roughly 1%) of people provide incorrect employment details but nearly 25% state a significantly higher income than their actual verifiable income.
- Secondly, we do not use sensitive information such as social security number, email, phone and the experience level for verifying income. This makes our approach broadly applicable since it can be applied in scenarios where the users are averse to giving out such private information. Of course, this information could help improve our models and in the future we would like to explore the possibility of using private information and even previous employment information to improve our models.

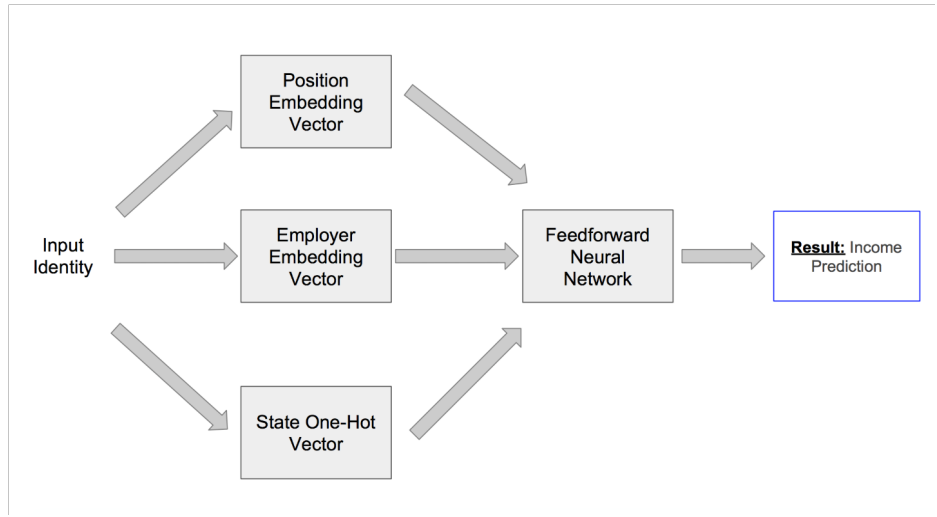

**Fig. 1.** Flowchart for internal model

## 2.2. Overall Approach

As described in the previous section, we want to determine whether the stated income is accurate for a user with the given identity and employment details. We cast the problem as a regression problem and use both the input data (e.g. employer, position, location) and online sources (e.g. payscale.com, Government databases) to predict the true income. Our model processes input and external data separately and combines the result. Initially, we built an internal model using only the input data. Figure 1 shows the model architecture for the processing of input data. Then we experimented with online sources and saw that it significantly improved model performance. Figure 2 shows the system architecture to utilize external data. It consists of several parts including, the input canonicalizer, a query generator for web and database records, data extractor, employment matcher, and feature extractor. The combined model is presented in Figure 3. We describe the individual components in the following sections.

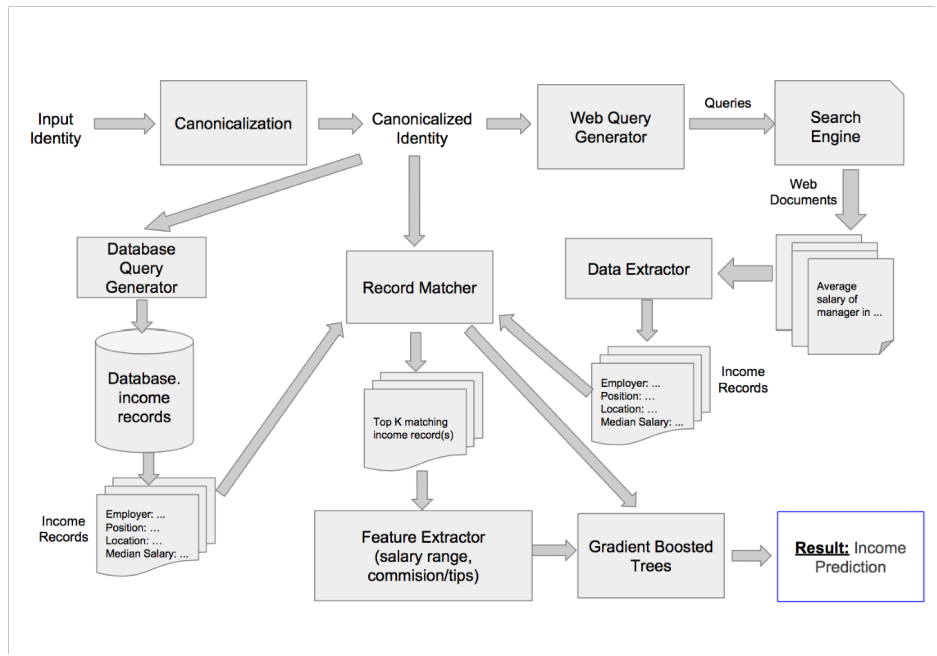

**Fig. 2.** Flowchart for external model

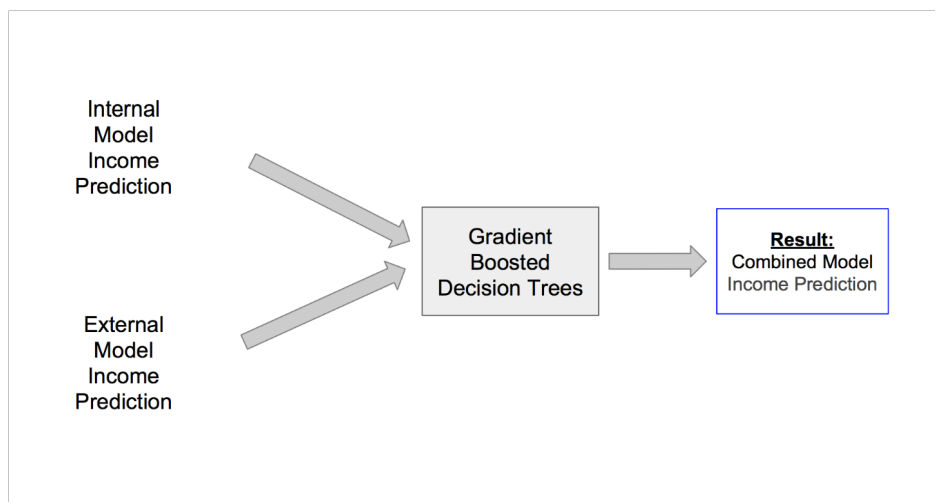

**Fig. 3.** Flowchart for combined model

## 2.3.    **Internal Model**

For the internal model we create the word embeddings trained on the job title and employer using word2vec model (Mikolov et al. 2013). The job title and the employer embeddings were further trained separately using publicly available position and user stated income data and publicly available employer and user stated income data. The original word embeddings were used as the embedding layer of a neural network which consisted of a LSTM layer, a dropout layer and a dense layer. LSTM recurrent neural networks are useful for sequential modeling tasks. LSTM units were developed by (Hochreiter and Schmidhuber 1997) to overcome gradient vanishing problem. They introduced the adaptive gating mechanism which decides the degree to which LSTM units keep the previous state and memorize the extracted features of the current data input.

We currently assume that user's location plays a role in determining the outcome. Currently we only incorporate information at the state-level by modeling user's state as a categorical variable. We use a one-hot encoding of the state as an indicator of user's location. In the future, we want to encode much finer-grained location information based on metropolitan statistical areas (MSAs) since they are widely used by Bureau of Labor Statistics.

The embedding dimensions were 300 each for employer and job title while the state encoding had a dimension of 50. This 650 dimension vector formed the input to a feed forward neural network. The network consisted of one hidden layer of size 200. We model the problem as a regression problem and our target is the income given by the input vector. We use Mean Absolute Error as the loss and otpimize the network parameters using gradient descent. We were able to get reasonable estimates of the income using the internal model. However, this did not use the wide range of sources available in the web.

### 2.4.  External Sources Overview

Recently, there has been a growing number of online sources collecting and disseminating income information (e.g. payscale.com, salary.com). The main reason for this push has been the need for pay transparency and accountability from both private and public institutions. In some cases such as government sources point estimates of the salary and bonus are available (see Table 2). Other sources collect compensation details from individuals willing to report such information and expose only anonymized ranges for individual income components (see Table 3).

In this paper, we describe a system to exploit such resources in order to improve income verification.

### 2.5.  External Model

The main components of the External Model are:
- *Canonicalizer*: In a large number of cases we find different representations of the same employer, job title. There were also spelling errors in the user input. Hence, the first step was to normalize the input. Some examples of noisy input and their normalized form is in Table 4. The transformation was performed using a lookup table. The table was manually created and consisted of the most common examples found in the dataset. In the future, we plan to explore better string and similarity models.
- *Web Query Generator:* Our main source of income records is from the web. We use targeted web queries as show in Table 5 to get the records of interest. E.g. if we want salaries for software engineers in XYZ company, the query would be 'XYZ Company Software Engineer Salary'. While there would be records available for large companies, for the long tail of companies we would rely on generic queries such as 'Software Engineer Salary'.
- *Record Matcher*: In this step, we extract the identities from the candidate records generated from database income records and search engines. We then choose the five best matches. The extraction system depends on the type of record. For records from structured sources such as Feds Data Center and Payscale, the identities were extracted using xml paths. For unstructured sources such as web snippets and text, we used pattern-based information extraction to extract the identity.

From the source identity and input identity we create features such as name match, address match, employment match, and industry match. To compute the name match score we use the normalized edit distance between the names of the input and the source. We also add a penalty for middle name dissimilarities (E.g. James Ryan Smith is not equal to James S Smith). We compute the address match by considering string similarity metrics for city, street match, and county, and exact matches for zipcode and country. We bucketed the matches into different confidence levels based on different points on the precision-recall curve. To compute the employment match we consider the cosine distance between the employer strings and the position strings.

We used a decision tree to match record pairs. In order to train our record matcher we annotated thousand input and source identity pairs. We then used different thresholds to bucket the model scores and assigned a confidence value to each bucket. E.g. A score of above 0.8 is a high match.

- *Feature Extractor*: We rank the sources according the match score and select the top 5 sources. We extract the base salary median, base salary low, base salary high, total compensation median, total compensation low and total compensation high for each of these sources. The total compensation consists of bonus, stock awards and profit sharing. In many cases only a subset of the salary attributes existed. We aggregated the employments by industry and found the industry wide ratio of each salary value to the other values. In cases of missing values, we filled it by multiplying the known salary attribute with the industry wide ratio. If none of the attributed existed, we discarded the source. We then used ratio of the attribute and the stated income as a feature. Each source generated 6 features. We also added the match score to the feature set. For each user, we had an feature dimension of 35.
- *Model*: We used the features from above and stacked them with the income pre-
diction from the input model and passed it to the gradient boosted decision tree as the final model.

## 3. Experiments

### 3.1. Datasets

We evaluated our model on two income datasets. The statistics of the dataset are provided in Table 6.

**Client Dataset** This dataset was obtained from one of the largest peer-to-peer lending companies. The input information includes names, home address, date of birth, employer name and job title. The client manually verified the salary information using the following methods:

- Requiring the loan applicants to submit documents such as paystubs, W-2 forms, or other tax records that verify the income stated in their loan request.
- Electronically checking their income data through a third party provider.
- Verifying the income with employer.

**H1-B Dataset** This dataset was obtained from the United States Department of Labor website[10] . The input information includes employer name, job title and address. The Department of Labor has been disclosing information about the H1- B visa class from 2008. This dataset was created from disclosure data as of 2016 and is only used to show the income prediction results since it does not contain a stated income as input.

**Table 2.** Example of Government source with fake names

| Name | Salary | Bonus | Agency | Location | Occupation | Year |
|---|---|---|---|---|---|---|
| James Bond | $73,482 | $0 | Department of Homeland Security | SELLS | Medical Technologist | 2016 |
| Harry Potter | $84,443 | $10000 | Department of Agriculture | AMES | General Engineering | 2016 |

**Table 3.** Example of information found online for software engineers of a random company

| | Mean | Min | Max |
|---|---|---|---|
| Base Salary | 150,000 | 90,000 | 234,000 |
| Total Compensation | 265,000 | 90,000 | 1,000,000 |
| Cash Bonus | 38,000 | 10,000 | 500,000 |
| Stocks | 100,000 | 10,000 | 800,000 |

**Table 4.** Canonicalization examples

| Input | Mean |
|---|---|
| U.S.P.S, U.S. Postal Service | United States Postal Service |
| GE, G.E | General Electric |
| Acc. Manager | Account Manager |
| Sr. Manager, Snr. Manager | Senior Manager |

### 3.2. Baselines

We compare our model with several methods such as neural networks and gradient boosted trees (GBT) with both input and external sources with different types of preprocessing. The model architectures and hyperparameters are the chosen based on the minimization of cross-validation error.

**Bag of Words (BOW) with Gradient Boosted Trees.** The top 200 words based on frequency are each selected from job title and employer name and the word counts are used as a feature. The city and state state are treated as categorical variables. The total input dimension is 402.

   The model is a gradient boosted tree with a linear regression objective, a max depth of 5, and an initial learning rate of 0.01.

**Mean Word Vectors with Feed-forward network.** The mean 300 dimension word2vec vectors (Mikolov et al. 2013) for job title and employer name trained on the input data are used as the feature set. The feature state is used as a one-hot encoded vector. The total input dimension is 650.

The model is a feed-forward network with two hidden layers of size 300 and 100 respectively.

**Mean Word Vectors trained on external data with Feedforward network.** The word2vec vectors are trained on the input data and external data from Lending Club's loan application dataset [11] . The data contains around 850,000 examples of job titles and 42,000 examples of employer names. This will generalize the word vectors on a larger dataset. The feature state is used as a one-hot encoded vector. The total input dimension is 650.

The model is a feed-forward network with two hidden layers of size 300 and 100 respectively.

**Tuned Mean Word Vectors trained on external data with Feed-forward network.** The word vectors are trained on the input and external data as in the previous model. The job title vectors are further trained by using it as an embedding in a deep neural network model predicting the user income in the Lending Club's loan application dataset mentioned above. The model has an embedding layer, a LSTM layer (128 dimensions), a dropout layer (rate of 0.5), and a hidden layer (200 dimensions). The model is trained on 520,000 examples and validated on 58,000 examples. The employer name vectors are trained in the same way as the previous model. The mean employer name, job title word vectors and the one-hot encoded state feature form an input dimension of 650.

The model is a feed-forward network with a hidden layer of size 200.

**External data with Gradient Boosted Trees.** The 5 best external sources with their salary low, median, high, and total compensation low, median, high are taken along with their match scores. The input dimension is 35.

The model is a gradient boosted tree with a linear regression objective, a max depth of 5, and an initial learning rate of 0.003.

### 3.3. Results

The experimental results on the datasets for the task of income prediction are presented in Table-7. The three metrics which we evaluate our model on are Cross Validation Mean Absolute Error (CV MAE), Test Set Mean Absolute Error (Test Set MAE) and the Test Set Mean Relative Error (Test Set MRE). The results show that the combined model performs the best on all metrics on both datasets with the exception of the MRE of the Client dataset. It outperforms the next best model by 2.8% on the Client dataset and 6% on the H1-B dataset.

From Table 7 we also observe that the performance of the model using only external data falls in the H1-B dataset where it is the fourth best model as compared to being the second best model in the client dataset. We hypothesize this is due to the lack of names and phone numbers in the input identity. This results in poor matches with database records. Despite this it is able to outperform the baseline model by 7.5%.

We also show the results of various models for the task of income verification in Table 8. This experiment was only performed on the Client dataset. Similar to the income prediction results, the combined model performs the best on this task as well.

**Table 5.** Example of search engine queries

| Query Template | Example |
| --- | --- |
| <Employer> <Job Title> Salary | XYZ Company Software Engineer Salary |
| <Job Title> Salary | Software Engineer Salary |
| <Industry> <Job Title> Salary | Travel Software Engineer Salary |

**Table 6.** Dataset Statistics

| Dataset | Size | Mean | Stddev | Skew |
|---|---|---|---|---|
| Client Train | 3108 | 77571.760 | 57979.323 | 5.694 |
| Client Test | 1037 | 78930.494 | 52488.707 | 3.576 |
| H1-B Train | 7500 | 91411.177 | 46969.135 | 1.878 |
| H1-B Test | 2500 | 90648.067 | 44355.742 | 1.407 |

### 3.4.    Ablation studies

We perform a number of experiments to study the impact of the number of sources, and the importance of each feature in the internal and external models. Our first experiment was to observe the variation in Test Set MAE with the number of web sources used as a feature. Figure 4 presents the graph of how the Test Set MAE varies with the number of sources. The data for the graph is presented in Table 9. The mean absolute error decreases while we increase the number of sources and then stabilizes after four sources. This makes intuitive sense because we expect the improvement from adding more sources to reduce after a certain threshold.

The second experiment we performed was to identify the importance of each feature in the input identity. This was evaluated by running the BOW + GBT model by removing one feature at a time. The results are shown in Table 10. We observe that removing the job title feature leads to the maximum increase in the Test Set MAE. One surprising result here is that removing the Employer Name does not degrade the model's performance significantly.

The third experiment we performed was to identify the impact of the different features which we extract from web and database records. The different features include low, median, high salary and total compensation. We remove each feature group (i.e. one of low, median and high) and retrain the External model. The results are presented in Table 11. We notice that removing the median leads to the maximum degradation in performance while there is no significant change on removing low and high. We hypothesize this is because most people earn close to the median and the median feature has the largest coverage in our dataset.

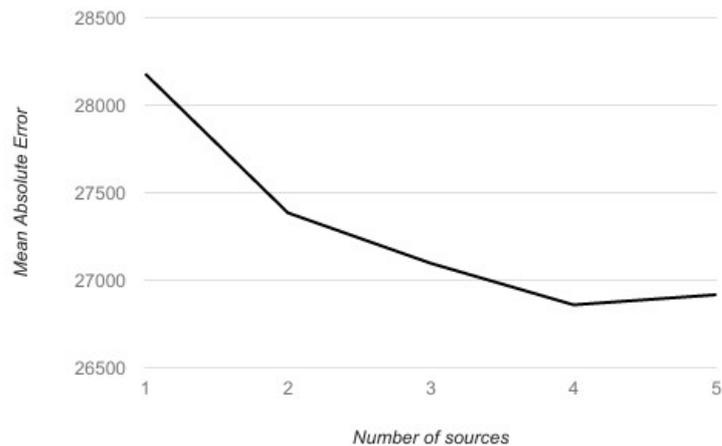

**Fig. 3.** Graph of the Test Set MAE vs. number of sources

## 4.     Related Work

(Kohavi 1996) used a hybrid of Naive Bayes and Decision Tree to predict whether a person makes over $50K a year in the Census Income dataset. The census dataset consists of categorical variables like education (e.g. Bachelors, Some-college, High school graduate), relationship (e.g. Wife, Own-child, Husband, Not-in-family) and continuous values like hours-per-week, age, capital gains. This dataset was collected from the 1994 Census. (Lazar 2004) applied Support Vector Machines on the same dataset to achieve superior results. However, in many cases banks and lending companies do not have access to all these features and need to build alternate approaches to model an individual's income.

Some of the recent work in predicting salaries has been motivated by the Kaggle's Adzuna Job Salary Prediction competition. The dataset provides 240,000 UK job postings each of which consist of title, job description, location, contract details, company as features. The goal of this competition is to model income based on the details available in the posting. There are no details about the individual's identity. (Li, Liu, and Zhou 2013) compared models such as Support Vector Regression, Linear Regression, K Nearest Neighbors Regression and Random Forest Regression and concluded that Random Forest works the best. (Jackman and Reid 2013) modeled the unstructured text fields as unigram bag of words and compared methods such as Maximum Likelihood, Dropout Neural Network and Random Forest. They reached the same conclusion that Random Forests outperform the other models.

(Preotiuc-Pietro et al. 2015) conducted a study to predict the income of social media users from their online behavior. They used profile features, psycho-demographic features, inferred emotion features, and text features. They discovered that emotional content, gender, race and education level have correlation with income. They were also able to highlight topics that distinguish users by income, such as politics, technology topics and swear words.

(Kibekbaev and Duman 2016) predict incomes of bank customers. Their work benchmarks various regression algorithms on five datasets provided by Turkish banks. However, the paper does not mention the features used and hence, it is difficult to compare their results against ours.

Recently, (Chen et al. 2018) proposed an approach to infer salary insights in the LinkedIn graph when there are limited or no user-reported salaries available for a given company. Their approach learns an embedding vector for each company based on employees transitioning between different companies and then applies a Bayesian statistical model to infer salary ranges based on similar companies in the embedding space. Our approach can definitely leverage some of the techniques used in this paper. However, a key advantage of our approach is that we are able to make accurate predictions without having access to the large amount of information within the LinkedIn graph.

**Table 7.** Experiment results of various models for the income prediction task (BOW = Bag of Words, GBT = Gradient Boosted Trees, WV = Word vectors, NN = Feed-forward neural network)

| Model | Client Dataset | | | H1-B dataset | | |
|---|---|---|---|---|---|---|
| | CV MAE | Test Set MAE | Test Set MRE | CV MAE | Test Set MAE | Test Set MRE |
| BOW + GBT | 31222.990 | 30126.355 | 0.425 | 23848.233 | 22611.401 | 0.251 |
| Mean WV + NN | 31536.852 | 30492.682 | 0.407 | 22812.092 | 21718.642 | 0.248 |
| External Mean WV + NN | 29834.140 | 28606.042 | 0.378 | 20055.863 | 18744.684 | 0.216 |
| Tuned Mean WV + NN | 30021.129 | 27917.228 | **0.361** | 19961.718 | 18591.964 | 0.212 |
| External data + GBT | 28402.790 | 26858.992 | 0.389 | 22639.001 | 20911.111 | 0.254 |
| Combined + GBT | **27710.378** | **26100.232** | 0.375 | **19000.542** | **17466.239** | **0.207** |

**Table 8.** Results of various models for the income verification task on the Client dataset

| Model | Precision | Recall | F1 score |
|---|---|---|---|
| BOW + GBT | 0.8196 | 0.6218 | 0.7072 |
| Mean WV + NN | 0.8328 | 0.5945 | 0.6938 |
| External Mean WV + NN | 0.8420 | 0.5833 | 0.6892 |
| Tuned Mean WV + NN | **0.8490** | 0.5945 | 0.6993 |
| External data + GBT | 0.8230 | **0.6766** | **0.7427** |

| Combined + GBT | 0.8230 | **0.6766** | 0.7427 |

**Table 9.** Results of external data model by the number of sources

| # sources | CV MAE | Test Set MAE | Test Set MRE |
|---|---|---|---|
| 1 | 28800.476 | 28179.456 | 0.422 |
| 2 | 28529.557 | 27384.617 | 0.393 |
| 3 | 28408.582 | 27096.461 | 0.394 |
| 4 | 28402.790 | **26858.992** | **0.389** |
| 5 | **28401.768** | 26916.694 | 0.390 |

**Table 10.** Results of BOW + GBT by removing one feature at a time

| Features | CV MAE | Test Set MAE | Test Set MRE |
|---|---|---|---|
| All features | 31222.990 | 30126.355 | 0.425 |
| - Job Title | **33651.446** | **33892.304** | **0.515** |
| - Employer Name | 31434.002 | 30000.765 | 0.453 |
| - State | 31500.645 | 29873.344 | 0.451 |
| - City | 31503.394 | 30121.302 | 0.454 |

**Table 11.** Results of External model by removing one feature group at a time

| Features | CV MAE | Test Set MAE | Test Set MRE |
|---|---|---|---|
| All features | 28401.768 | 26916.694 | 0.390 |
| - Low | 28566.334 | 26997.413 | 0.389 |
| - Median | **28810.107** | **27972.642** | **0.404** |
| - High | 28385.222 | 27026.328 | 0.394 |

## 5. Conclusions

In this paper, we introduce the problem of income verification and propose a model which used internal features such as position, employer, location along with external features such as salary information in web documents and government database records. Our experiments showed that the combined model performed better than all the other models. We found that job title is the most important input feature and salary median is the most important external feature. We also found that the model MAE depends on the number of external sources used.